\documentclass[prb,twocolumn,aps]{revtex4}
\usepackage{graphicx}
\usepackage{dcolumn}
\usepackage{bm}
\usepackage{hyperref}

\newcommand{\beq}{\begin{eqnarray}}
\newcommand{\eeq}{\end{eqnarray}}

\begin{document}\title{ Charge 2e Boson Underlies Two - Fluid Model of
  the Pseudogap  in Cuprate Superconductors}
\author{\small Shiladitya Chakraborty }
\author{Philip Phillips}
\affiliation{Department of Physics,
University of Illinois
1110 W. Green Street, Urbana, IL 61801, U.S.A.}
\date{\today}
 
\begin{abstract}
    Starting from the effective low energy theory of a doped Mott
    insulator\cite{ftm1,ftm2,ftm3}, we show that the effective carrier density in the underdoped regime agrees with a two - fluid description. Namely, it has distinct temperature independent  and thermally activated components. We identify the thermally activated component as the bound state of a hole and a charge 2e boson,
which occurs naturally in the effective theory. The thermally
activated unbinding of this state leads to the strange metal and
subsequent $T-$linear resistivity. We find that the doping dependence of the binding energy is in excellent agreement with the experimentally determined pseudogap energy scale in cuprate superconductors.
\end {abstract}

\maketitle
 
The normal state of the high-$T_c$ copper oxide superconductors exhibits a variety of
anomalous features in the underdoped regime  which any successful
theory of these materials must explain.  Central to the
 exotica of the underdoped cuprates are the
pseudogap\cite{timusk,alloul} and strange metal phases.  These phases
are closely linked because once the suppression of the density of
states at the chemical potential, a key experimental signature of the
pseudogap, ceases at some critical temperature, $T^\ast$, a metallic
state ensues.  Such behavior is suggestive of a localized, or more
properly, a `bound' electronic state that is liberated at $T^*$. While
the upturn\cite{boebinger,hussey} of the resistivity at low
temperatures is consistent with
this bound state scenario or charge localization\cite{choy,weng,schmalian,kampf} a more direct signature is the activated
temperature dependence\cite{nishikawa, ando, ono} of the Hall
coefficient.  In a Fermi liquid, the inverse of the Hall coefficient
is a measure of the carrier density which of course is independent of temperature.  However,
in the underdoped cuprates, the inverse of the Hall coefficient is
strongly temperature dependent\cite{nishikawa, ando, ono}. Gor'kov and Teitel'baum\cite{gorkov}
observed remarkably that the charge carrier concentration, n$_{\rm Hall}$,
extracted from the inverse of the Hall coefficient in
La$_{2-x}$Sr$_x$CuO$_4$ (LSCO) obeys an empirical formula,
\beq\label{n} 
n_{\rm Hall}(x,T)= n_{0}(x) + n_{1}(x)\exp(-\Delta(x)/T),
\eeq
appropriate or a two-component or two-fluid system.  One of the
components is independent of temperature, n$_0(x)$ ($x$ the doping level) while the other is
strongly temperature dependent, $n_{1}(x)\exp(-\Delta(x,T))$.  
The key observation here is that the temperature dependence in
$n_{\rm Hall}$ is carried entirely within $\Delta(x,T)$ which defines a
characteristic activation energy scale for the system.  Gor'kov and
Teitel`baum's\cite{gorkov} analysis suggests that the activation energy is set by the pseudogap energy
scale.  Consequently, the bound component should be liberated beyond
the $T^*$ scale for the onset of the pseudogap.  Should
$n_{\rm Hall}$ be an accurate representation of the effective charge
carrier concentration in the cuprates, the above observation indicates
that the underdoped or pseudogap phase necessitates a two-fluid
description, which has been championed\cite{pines} recently to explain NMR, inelastic neutron scattering
and thermodynamic measurements on these systems.  Nonetheless, the microscopic origin
of the two fluids has not been advanced. That is, there is no
microscopic prescription for the precise nature of the propagating
degrees of freedom that underlie the temperature dependence of $n_{\rm
  Hall}$.  For example, Gor`kov and
Teitel`baum\cite{gorkov} attributed the unbinding of the localized
charges above $T^*$ to excitations from van Hove singularities at the
bottom of the band  
up to the chemical potential. 

By contrast, our explanation of the the two fluids relies entirely on the
strong correlations of a doped Mott insulator, that is, Mottness. 
Here we show that the exact low-energy theory of a doped Mott
insulator\cite{ftm1,ftm2,ftm3} described by the Hubbard model
naturally resolves the two-component conundrum in the cuprates.  The
propagating degrees of freedom that constitute the two fluids are the standard projected electron in
the lower Hubbard band and a bound composite excitation composed of
a charge 2e boson and a hole.  It is the unbinding of the latter that
gives rise to the strange metal regime. The binding energy is found to be in excellent agreement with experimental values for the pseudogap energy scale.

We review some of the key features of the our effective low
energy theory of Mottness, the complete details of which have been worked out
elsewhere\cite{ftm1, ftm2, ftm3,ftm4}. Our starting point is the usual
Hubbard model
\beq
H_{\rm Hubb}&=&-t\sum_{i,j,\sigma} g_{ij} c^\dagger_{i,\sigma}c_{j,\sigma}+U\sum_{i,\sigma} c^\dagger_{i,\uparrow}c^\dagger_{i,\downarrow}c_{i,\downarrow}c_{i,\uparrow}
\eeq
where $i,j$ label lattice sites, $g_{ij}$ is equal to one if $i,j$ are
nearest neighbours, $c_{i\sigma}$ annihilates an electron with spin
$\sigma$ on lattice site $i$, $t$ is the nearest-neighbour hopping
matrix element and $U$ the energy cost when two electrons doubly
occupy the same site. The cuprates live in the strongly coupled regime
in which the interactions dominate as $t\approx 0.5$eV and $U=4$eV.  A
low-energy effective action is then obtained by integrating out the
physics on the $U$-scale.  Because double occupancy occurs in the
ground state, integrating out the $U$-scale physics is not equivalent
to integrating out double occupancy.  We solved this problem by
extending the Hilbert space to include a new fermionic oscillator which
represents the creation or annihilation of double occpancy only when a
constraint is solved.  The new fermionic oscillator enters the
action with a mass of $U$ and hence represents the high-energy scale,
which must be integrated out to generate the low-energy action. The corresponding
low-energy theory contains new degrees of freedom, namely a charge 2e
boson, denoted by $\varphi_i$, that are absent in the original model
and  are not made out of the elemental excitations. $\varphi_i$ enters the
theory initially as the Lagrange multiplier to maintain the constraint
that in the extended Hilbert space the heavy fermionic field
represents the creation of double occupancy.   To leading order in $t/U$ , the effective Hamiltonian 
\beq
H_{\rm eff} &=& -t \displaystyle \sum_{i,j,\sigma}g_{ij}\alpha_{ij\sigma}c^\dagger_{i\sigma} c_{j\sigma}
-\frac{t^2}{U}\displaystyle \sum_{j}b^\dagger_j b_j 
-\frac{t^2}{U}\displaystyle \sum_{j}\varphi^\dagger_i \varphi_i\nonumber\\
&-& t\displaystyle \sum_j \varphi^\dagger_j c_{j\uparrow} c_{j\downarrow} -\frac{t^2}{U} \displaystyle \sum_i \varphi^\dagger_i b_i + h.c.
\eeq
contains the $t-J$ model (the first two terms) and new interactions with the charge 2e
boson, $\varphi_i$ which represent mixing with the sectors with
varying numbers of doubly occupied sites.  Here $b_i = \displaystyle
\sum_j b_{ij} =\displaystyle \sum_{j\sigma}g_{ij}c_{j,\sigma}V_\sigma
c_{i ,-\sigma}$ with $V_\uparrow = -V_\downarrow = 1$.   The $|b_i|^2$
term contains the spin-spin interaction as well as the three-site
hopping term.  A gradient expansion of this term shows that the
spin-spin term scales as $a^4$, $a$ the lattice constant, whereas the
terms linear in $b$ are proportional to $a^2$.  Hence, relative to the
terms linear in $b$, the $|b|^2$ term is irrelevant.  Our key
contention which has been worked out extensively for the cuprates\cite{ftm1,ftm3,ftm4} is that as far as the charge degrees of freedom are
concerned, it is the interactions with the $\varphi_i$ sector that
determine the propagating degrees of freedom, not the dynamics arising
from the spin-spin term.  In particular, we show that it is the
$\varphi$ terms that give rise to a gap in the single-particle
electron spectrum.  As this gap is on the order of $t$, any
contribution from the spin-spin term will be subdominant. 

It is important to realise that once the heavy field is integrated
out, the Hilbert space reverts back that of the Hubbard model.
Further,
as $\varphi_i$ has no bare dynamics and $\varphi$ has no Fock space of its own, its
only contribution will be to create bound states within the Hilbert
space of the Hubbard model.  This can be seen initially by considering
how the electron operator transforms\cite{ftm1,ftm2,ftm3,ftm4}
\beq\label{cop}
c^\dagger_{i,\sigma}&\rightarrow&(1-n_{i,-\sigma})c_{i,\sigma}^\dagger
+ V_\sigma \frac{t}{U} b_i c_{i,-\sigma}\nonumber\\
&+& V_\sigma \frac{t}{U}\varphi_i^\dagger c_{i,-\sigma}
\eeq
 upon the integration of the high
energy scale.  The first two terms represent the standard electron
operator in the lower Hubbard band dressed with spin fluctuations.
However, the last term represents the correction due to dynamical
spectral weight transfer\cite{meinders}, that is the mixing with
doubly occupied sites.  All such processes are mediated by $\varphi_i$
which represents a collective charge 2e mode arising from the dynamics
of double occupancy.  The quantity $\varphi_i^\dagger c_{i,-\sigma}$
represents a bound complex with a non-propagating local degree of
freedom.  It is from this term that the bound state dynamics emerges.
Roughly, the two-fluid emerge from the fact that an electron at low
energies is a linear superposition of an essentially free part, the
first two terms in Eq. (\ref{cop}) and the last term in
Eq. (\ref{cop}) from which the bound -state\cite{ftm3,ftm4} dynamics (that is,
pseudogap physics) emerges. As a result of $\varphi_i$, the  conserved charge is no longer just the total number of electrons but 
{\beq 
Q = \displaystyle \sum_{i\sigma}c^\dagger_{i\sigma}c_{i\sigma} + \displaystyle 2\sum_{i}\varphi^\dagger_{i}\varphi_{i}
\eeq}

In  order to obtain the electron Green function for the effective
Hamiltonian, we treat
$\varphi_i$ as spatially independent, owing to a lack of gradient terms
in that field in the Hamiltonian. To keep track of the dependence on
$\varphi_i$ it is helpful to introduce the rescaling
$\varphi_i\rightarrow s\varphi_i$.  The electron Green function is then written as a path integral over the $\varphi$ fields as
\beq
\label{Green}
G(\bf{k},\omega)&=&\frac{1}{Z}
\int[D\varphi^*][D\varphi]FT(\int[dc^*_i][dc_i]c_i(t)c^*_i(0)\nonumber\\
&\times&\exp{\left(-\int L(c,\varphi) dt\right)})
\eeq
where the effective Lagrangian $L$ is expressed in a diagonalized form
\beq
L&=& \displaystyle \sum_{k\sigma}\gamma^*_{k
  \sigma}\dot{\gamma}_{k\sigma}+\displaystyle \sum_k (E_0 + E_k -
\lambda_k) + \displaystyle \sum_{k \sigma}
\lambda_k \gamma^*_k \gamma_k,\nonumber\\
\label{leff}
\eeq
where the $\gamma_{k\sigma}$ are the Boguliubov quasiparticles and are given by
\beq
\gamma^*_{k\uparrow}= \cos\theta_k c^*_{k\uparrow} +\sin\theta_k c_{-k\downarrow}
\eeq
\beq
\gamma_{k\downarrow}=-\sin\theta_k c^*_{k\uparrow} + \cos\theta_k c_{-k\downarrow}
\eeq
where $\cos^2\theta_k =\frac{1}{2}(1 + \frac{E_k}{\lambda_k})$, $\alpha_k = 2(\cos k_x + \cos k_y)$ , $E_0 = (-2\mu + \frac{s^2}{U})\varphi^*\varphi
$,$E_k = -g_t t\alpha_k -\mu$,$\lambda_k = \sqrt{E^2_k
  +\Delta^2_k}$, the gap is proportional to $s$, $\Delta_k =
s\varphi^*(1-\frac{2t}{U}\alpha_k)$, and hence vanishes when $\varphi$
is absent and
$g_t = \frac{2\delta}{1 + \delta}$, $\delta = 1 -n$.  The $g_t$ term
originates from the correlated hopping term,
$(1-n_{i\bar\sigma})c_{i\sigma}^\dagger c_{j\sigma}(1-n_{j\bar\sigma})$.
The
The $\gamma_{k\sigma}$'s play the role of the fundamental low energy
degrees of freedom in a doped Mott insulator. That is, they are the
natural propagating charge degrees of freedom.  Note they depend in a
complicated way on the the $\varphi_i$ field and consequently are
heavily mixed with the doubly occupied sector. Starting from Eq. (\ref{leff}), we integrate over the fermions in Eq. (\ref{Green}) to obtain,
\begin{widetext}
\beq\label{eqG}
G(k,\omega) = \frac{1}{Z}\int [D\varphi^*] [D\varphi] G(k,\omega, \varphi) \exp^{-\sum_k (E_0 + E_k - \lambda_k -\frac{2}{\beta} \ln(1+e^{-\beta\lambda_k}) )}
\eeq
\end{widetext}
where
\beq\label{gfinal}
G(k,\omega,\varphi)=\frac{\sin^2\theta_k[\varphi]}{\omega+\lambda_k[\varphi]} + \frac{\cos^2\theta_k[\varphi]}{\omega-\lambda_k[\varphi]}
\eeq
is the exact Green function corresponding to the Lagrangian, Eq. (\ref{leff}),
which has a two-branch structure, corresponding to the bare electrons
and the coupled holon-doublon state respectively. The role of the
$\varphi$ field, which determines the weight of the second branch, is
vital to our understanding of the properties of Mott systems, as
was demonstrated previously\cite{ftm3,ftm4}. It is trivial  to see that in the
limit of vanishing $s$ (no  $\varphi$ field), the $\gamma_{k\sigma}$'s
reduce to the bare electron operators $c_k$ and the first term in
Eq.(\ref{gfinal}) vanishes.   The two-fluid nature of the response
stems from this fact of the theory.  Namely, the first term
contributes only when $\varphi\ne 0$ and the second when $\varphi=0$.
These contributions correspond to the dynamical and static components of the
spectral weight, respectively.
 
We obtained the Green function $G(\bf{k},\omega)$ by a numerical integration of Eq.(\ref{eqG}) over the $\varphi$ field. The Hall coefficient $R_H$ was computed from the spectral function $A(\bf{k},\omega)$ using the Kubo formula\cite{jarrell}
\beq
R_H = \sigma_{\rm xy}/\sigma_{\rm xx}^2,
\eeq
where
\beq
\sigma_{\rm xy}& =& \frac{2\pi^{2}|e|^{3}aB}{3\hbar^{2}}\int d\omega
(\frac{\partial f(\omega)}{\partial \omega})\frac{1}{N} \displaystyle
\sum_{\bf{k}}(\frac{\partial \epsilon_{\bf{k}}}{\partial{k_x}})^2
\nonumber\\
&\times&\frac{\partial^2 \epsilon_{\bf{k}}}{\partial {k_y}^2} A(\bf{k},\omega)^3 
\eeq
and
\beq
\sigma_{\rm xx} = \frac{\pi e^2 }{2 \hbar a} \int d\omega (-\frac{\partial f(\omega)}{\partial \omega})\frac{1}{N} \displaystyle \sum_{\bf{k}}(\frac{\partial \epsilon_{\bf{k}}}{\partial{k_x}})^2 A(\bf{k},\omega)^2
\eeq
with $\sigma_{\rm xx}$ and $\sigma_{\rm xy}$ the diagonal and off-diagonal
components of the conductivity tensor respectively, $f(\omega)$ is the
Fermi distribution function, and $B$ is the normal component of the
external magnetic field. The effective charge carrier density $n_{\rm
  Hall}$ is then obtained using the relation $R_H = -1/(n_{\rm Hall} e)$.
 
\begin{figure}
\includegraphics[width = 8.5cm]{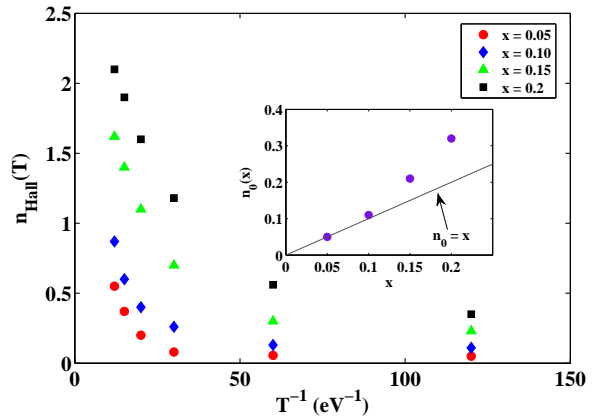}
\caption{n$_{\rm Hall}$ plotted as a function of inverse temperature for
  four different values of hole doping $x$: 1) solid circles,
  $x=0.05$, 2) diamonds, $x=0.10$, 3) triangles, $x=0.15$, and 4)
  squares, $x=0.2$. The inset shows the
  temperature independent part of the carrier density as a function of
  $x$.  Note it exceeds the nominal doping level indicated by the
  straight line. }
\label{fig1}
\end{figure}

Fig.\ref{fig1} shows a set of plots of $n_{\rm Hall}$ as a
function of the inverse temperature, each corresponding to a different
value of hole-doping, $x$, in the underdoped regime ($x$ ranging from
0.05 to 0.20). The plots fit remarkably well to an exponentially
decaying form.  In other words, the computed charge carrier density
within the charge 2e boson theory of a doped Mott insulator
agrees well with the form given in Eq. (\ref{n}) proposed by Gor'kov
and Teitel'baum\cite{gorkov}. The inset shows the
temperature-independent part of the charge density as a function of
$x$.  This quantity exceeds the nominal doping level.  This deviation
is expected as the Hall coefficient is expected to change sign around
$x=0.3$\cite{hall} in hole-doped samples.

\begin{figure}
\includegraphics[width = 8.5cm]{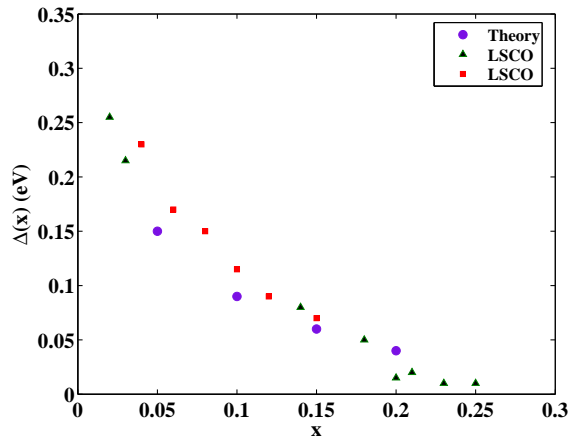}
\caption{$\Delta(x)$ (solid circles) obtained from fitting the plots in
  Fig.(\ref{fig1}) to Eq.(~\ref{n}) plotted as a function of hole
  doping $x$. The experimental values are also shown for LSCO: solid 
triangles\cite{ando,ono,delta1} and squares\cite{nishikawa}  The
  excellent agreement indicates that the bound component contributing
  to the charge density does in fact give rise to the pseudogap.}
\label{fig2}
\end{figure}

\begin{figure}
\includegraphics[width = 9.0cm]{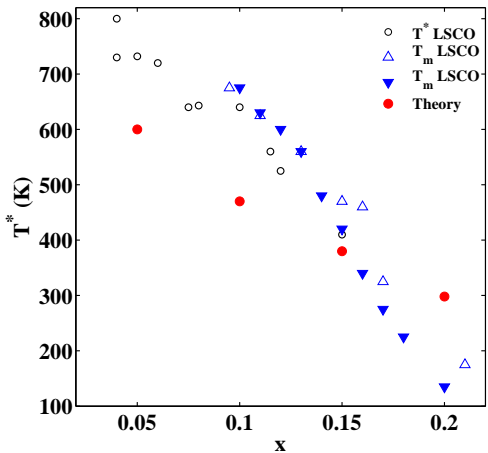}
\caption{$T^*(x))$ (solid circles) obtained from  Eq.(~\ref{T*})
  plotted as a function of hole doping $x$. The experimental data were
  gleaned from the following: open circles are from Ref.\cite{timusk}$T^*$,  open triangles ($T_m$) from Ref.\cite{23}, and closed triangles ($T_m$) from Ref.\cite{24}. }
\label{fig3}
\end{figure}

 The `binding energy', $\Delta(x)$, was extracted for each doping and plotted in
 Fig.(\ref{fig2}) using Eq. (\ref{n}).  Shown here also are the values
 for the experimentally determined pseudogap energy for LSCO\cite{ando,ono,delta1}.  The
 magnitude of $\Delta(x)$ falls with increasing hole doping as is seen
 experimentally and hence is consistent with its interpretation, even quantitatively, as a
 measure of the pseudogap temperature $T^*$. A more accurate estimate
 of $T^*$,
 \beq
 \label{T*}
 T^*(x)\approx -\Delta(x)/\ln(x),
 \eeq
may be obtained from $\Delta(x)$, by equating the number of doped
carriers $x$ with that of the activated ones
$n_{1}(x)exp(-\Delta(x,T))$ as proposed by Gor'kov and Teitel'baum\cite{gorkov}.
Fig.(\ref{fig3}) shows a plot of $T^*$ as a function of $x$. This is
in reasonable agreement with the experimentally obtained estimates of
$T^*$\cite{timusk,23,24}.  

Ultimately it is not surprising that the pseudogap\cite{ftm1,ftm3} appears within
the charge 2e boson theory.  As mentioned previously, the charge 2e
boson is a local collective non-propagating mode that is restricted
to mediate electronic states within the Hilbert space of the Hubbard
model. The only option is that the boson binds to a hole to form a new
charge e state.  As an electron at low-energies (Eq. (\ref{cop})) is a  linear
superposition of the standard state in the lower Hubbard band and the
bound state mediated by the charge 2e boson, a two-fluid charge model
is a natural consequence.  
This further supports the
idea\cite{ftm3} that  the pseudogap temperature $T^*$
represents  the boundary between bound and unbound charge $2e$ bosons where the
binding energy to excite a boson vanishes, and T-linear resistivity
obtains\cite{ftm3,ftm4}.  The mechanism for $T-$linear resistivity is simple within
this model. Once the binding energy of the boson vanishes, bosons are
free to scatter off the electrons.  The resistivity of electrons
scattering off of bosons is well-known to scale linearly with temperature above the
energy to create a boson.  Hence, this mechanism is robust and should
persist to high temperatures.  Consequently, the charge 2e boson
theory offers a resolution of the pseudogap and the transition to the
strange metal regime of the cuprates.  

\acknowledgements P. Phillips thanks John Tranquada for the invaluable
suggestion
that our theory should be able to explain the two-fluid structure of
the effective charge density proposed by Gor'kov and Teitel'baum and the NSF DMR-0605769 for partially
funding this work.

 \end{document}